\documentclass[letterpaper, 12pt]{article}
\usepackage{latexsym}
\usepackage{amsmath}
\usepackage{amssymb}
\usepackage[latin1]{inputenc}
\usepackage{graphicx}
\def\CE{{\cal E}}
\def\CF{{\cal F}}

\def\CM{{\cal M}}

\def\CO{{\cal O}}

\newcommand{\ads}[1]{{\rm AdS}_{#1}}

\newcommand{\be}{\begin{equation}}
\newcommand{\ee}{\end{equation}}
\newcommand{\bea}{\begin{eqnarray}}
\newcommand{\eea}{\end{eqnarray}}

\newcommand{\bbibitem}[1]{\bibitem{#1}\marginpar{#1}}

\def\Label#1{\label{#1}%
  \smash{\hbox to0pt{\raise1ex\hbox{\tiny[#1]}\hss}}}
\def\noLabels{\let\Label=\label}
\def\nobbibitem{\let\bbibitem=\bibitem}

\begin{document}
\noLabels
\nobbibitem

\rightline{UPR-T-1142, hep-th/0602263}

 \vskip 1cm

\centerline{\Large \bf Integrability vs. Information Loss:  A
Simple Example} \vskip 1cm

\renewcommand{\thefootnote}{\fnsymbol{footnote}}
\centerline{{\bf Vijay Balasubramanian${}$\footnote{vijay@physics.upenn.edu},
Bart{\l}omiej Czech${}$\footnote{czech@sas.upenn.edu},
Klaus Larjo${}$\footnote{klarjo@physics.upenn.edu} and Joan Sim\'{o}n${}$\footnote{jsimon@bokchoy.hep.upenn.edu},
}}
\vskip .5cm
\centerline{\it David Rittenhouse Laboratories, University of Pennsylvania,}
\centerline{\it Philadelphia, PA 19104, U.S.A.}

\vskip 1cm
\setcounter{footnote}{0}
\renewcommand{\thefootnote}{\arabic{footnote}}
\begin{abstract}
The half-BPS sector of Yang-Mills theory with 16 supercharges is integrable: there is a set of commuting conserved charges, whose eigenvalues can completely identify a state.   We show that these charges can be measured in the dual gravitational description from  asymptotic multipole moments of the spacetime.   However, Planck scale measurements are required to separate the charges of different microstates.  Thus, semiclassical observers making coarse-grained measurements necessarily lose information about the underlying quantum state.
\end{abstract}

%\newpage
%\tableofcontents
\newpage 

\section{Introduction}
\Label{sec:intro}

%In statistical physics we are familiar with many microscopic
%states having an identical macroscopic description. A familiar
%example of this is the ideal gas law, which gives a universal
%macroscopic description of many distinct microscopic
%configurations of gas molecules. Lately this idea of a universal
%effective description has been also applied to gravitational
%settings \cite{vijayjoan, mathur}. In this paper we explore what
%sort of data an asymptotic can observer measure about the
%underlying microscopic state and what information is lost in the
%universal description.

The half-BPS sector of $SU(N)$ Yang-Mills theory with 16
supercharges contains a complete commuting set of $N$ conserved
charges $M_k$, and an eigenstate of these $M_k$ can be identified
from the eigenvalues. The gravitational duals of these states in
$\ads{5}$ have also been identified \cite{llm}, and it has been
argued that typical half-BPS states behave universally like an
extremal black hole in response to almost all probes
\cite{vijayjoan}.  This implies that information about the
underlying microstate is lost.     Here we show how this tension
between integrability and information loss is resolved.   We
demonstrate that angular moments of the gravitational solution
that can be read off from the asymptotic metric directly measure
the higher conserved charges of the underlying quantum microstate.     The low
moments have magnitudes large enough for  semiclassical
observation, but measuring the differences in these moments
between typical states requires Planck scale precision.  The high
moments vary strongly between typical states, but their magnitudes
are so small that Planck scale precision is again required to
measure them.  Thus, the coarse-grained semiclassical
gravitational description of the underlying exact quantum
microstate necessarily loses information.

\section{Half-BPS states}
\Label{sec:review}

\subsection{Description in gauge theory}
\Label{reviewgauge} $SU(N)$ Yang-Mills theory on $S^3 \times R$
with 16 supercharges has a spectrum of half-BPS states.  The
highest weight representatives in each BPS multiplet are created
by operators that are gauge invariant polynomials in the
zero-modes of a single adjoint scalar field. The Hamiltonian of
the gauge theory restricted to these states reduces to that of a
Hermitian matrix harmonic oscillator \cite{cjr,toymodel}.  This is
solved by going to the eigenvalue basis, and redefining fields to
arrive at a theory of $N$ free fermions in a harmonic potential.
A basis of  highest weight half-BPS states is completely specified
by a set of increasing integers ${\cal F} =\{f_1,\,f_2,\,\dots
,\,f_N\}$ that provide the excitation numbers of each individual
fermion $E_i = \hbar\left(f_i + \frac{1}{2}\right)$, $i=1,\dots
,N$.   Equivalently, we can reproduce the state by giving a set of
non-decreasing integers $\{r_i\}$, which measure the excitation of
any given fermion above the vacuum
\begin{equation}
r_i = \frac{E_i}{\hbar}-i+\frac{1}{2} \,. \Label{rdef}
\end{equation}
This data can be encoded  as a Young diagram in which the $i$-th row has length $r_i$.

The energies $\{ f_1,\cdots f_N \}$ completely specify these basis
states.  The states can also be specified in terms of the moments 
\begin{equation}
M_k = \sum_{i=1}^N f_i^k = {\rm Tr}(H_N^k/\hbar^k)  ~~~;~~~ k=0,\cdots N \, ,
\Label{momentdef}
\end{equation}
where $H_N$ is the Hamiltonian acting on the $N$ fermion Hilbert space with the zero point energy removed.   Manifestly, the $M_k$ are conserved charges of the system of fermions in a harmonic potential \cite{surya}.      The basis of states with fixed fermion excitation energies that was described above consists of eigenstates of the moment operators.  The individual excitation energies ${\cal F}$ in these eigenstates can always be reconstructed from the set of moments $\CM = \{M_0, M_1, \cdots M_N \}$.  To see this, construct the  characteristic polynomial
\begin{equation}
P = \det (x \, \textrm{I}_N - H_N/\hbar) = \prod_{i=1}^N (x-f_i) \, .
\Label{charpoly}
\end{equation}
This product can be expanded in symmetric products of the $f_i$ as
\begin{equation}
P  =  \sum_{p=0}^N (-1)^p \pi_p x^{N-p}   ~~~~;~~~~ \pi_k = \sum_{i_1 < i_2 < \cdots i_k} f_{i_1} f_{i_2} \cdots f_{i_k} \, .
\end{equation}
The $\pi_i$ are given recursively in
terms of the moments $M_k$ by the Newton--Girard formula
\begin{equation}
m \pi_m + \sum_{k=1}^m (-1)^k M_k \pi_{m-k} = 0.
\end{equation}
Thus, given a measurement of the moments $M_k$ for $k=1,\cdots N$ one can compute the symmetric products $\pi_i$ and from these determine the fermion excitations ${\cal F}$ that completely determine the basis BPS states as the roots of the characteristic polynomial (\ref{charpoly}).

Very heavy half-BPS states ($M_1 = {\rm Tr}(H_N/\hbar) \sim N^2$) have a dual description in terms of large-scale classical solutions of the maximally supersymmetric gravity in $\ads{5}$.   In \cite{vijayjoan} it was shown that almost all such eigenstates of the moment operators $\{ M_k \}$ lie very close to a certain ``typical state'' with a characteristic distribution of fermion energies.     In an ensemble of states in which the maximum excitation energy of any given fermion is bounded (so that $r_i \leq N_c \, \forall \, i$) the expected excitation energies of the fermions in the typical state are given by:
\begin{equation}
\langle r_i \rangle =  \sum_{j=0}^{i-1} {e^{-\beta(N-j) - \lambda} \over 1 - {e^{-\beta(N-j) - \lambda}}}
\Label{typicaldiscrete}
\end{equation}
where $\beta$ is an ``inverse temperature'' used to fix the total energy and $\lambda$ is a ``chemical potential'' fixing the maximum excitation of individual fermions.    The  semiclassical limit for this system is obtained by sending $\hbar \to 0$ with $N\hbar$ kept constant to fix the Fermi level.   In this limit, almost all half BPS states have fermion energy distributions that are small fluctuations around (\ref{typicaldiscrete}).   In this limit we can treat the fermion number $i$ and the excitation energies $r_i$ as continuous variables and relabel
\begin{equation}
i \to x ~{\rm and}~ r_i \to y(x) \, .
\Label{limit1}
\end{equation}
Any eigenstate of the moment operators that has a semiclassical limit can then be described in terms of such a function $y(x)$.  In particular, the typical state  (\ref{typicaldiscrete}) is summarized as a limit curve
\begin{equation}
C(N,N_C) \, e^{-\beta (N -x)} + D(N,N_c) \, e^{-\beta (y - N_c)} = 1 \, ,
\Label{limit2}
\end{equation}
where $C$ are $D$ are determined  by fixing the total energy and maximum excitation.    The maximum entropy results when $\beta \to 0$ in which case this limit curve simply reduces to
\begin{equation}
y = {N_c \over N } x
\Label{superstarlimit}
\end{equation}
Thus, in the large $N,N_c$ limit, almost all half-BPS states with bounded fermion excitations lie very close to the line (\ref{superstarlimit}).   Said otherwise, the {\it uniform} distribution over all sets of integers $r_1 \leq r_2 \leq \cdots r_N \leq N_c$ is strongly localized on partitions that lie close to the curve (\ref{superstarlimit}).  We will have use of this fact later.

These half-BPS states can also be described in terms of a distribution of fermions on the single particle harmonic oscillator phase space $(p,q)$ \cite{vijayjoan}.    Wigner \cite{wignerreview} described a distribution function that encodes the expectation value of Weyl (symmetric) ordered quantum observables,
\begin{equation}
\int dp \, dq \, W(p,q) \, f(p,q) = \langle \CO_W(f) \rangle \, ,
\end{equation}
where $\CO_W$ is the Weyl-ordered operator corresponding to the classical function $f$.  The Wigner distribution corresponding to a state $\mathcal{F}$ was computed in
\cite{vijayjoan} to be
\begin{equation}
W(r) = \frac{1}{\pi \hbar} e^{-\frac{r^2}{\hbar}} \sum_{f\in
\mathcal{F}} (-1)^f L_f\left(\frac{2r^2}{\hbar}\right),
\Label{eq:wigner}
\end{equation}
where $L_f(x)$ is a Laguerre polynomial and $r^2 = q^2 + p^2$. Since we are considering an eigenstate of the Hamiltonian, the corresponding phase  space distribution is rotationally invariant.    If a state is an eigenstate of the $M_k$, all of these moments can be extracted from the Wigner distribution.  Two other interesting distributions on phase space were identified in \cite{vijayjoan}.  The Husimi distribution
\begin{equation}
\textrm{Hu}(r) = \frac{1}{2\pi \hbar} e^{-\frac{r^2}{\hbar}}
\sum_{f\in \mathcal{F}} \frac{1}{f!}
\left(\frac{r^{2}}{\hbar}\right)^f \Label{husimi}
\end{equation}
arises by smoothing the Wigner distribution with a Gaussian kernel at the $\hbar$ scale, and computes
the expectation values of reverse normal ordered operators \cite{husimi}.      Finally, consider a state that is described in the semiclassical $\hbar \to 0$ limit by a limit curve $y(x)$ as in (\ref{limit1}, \ref{limit2}).   It was shown in \cite{vijayjoan} that a semiclassical observer of such a state, having access only to areas larger than $\hbar$ in phase space, interacts with the effective ``grayscale distribution''
\begin{equation}
g(r) = {1 \over 1 + dy/dx} \, .
\Label{grayscale}
\end{equation}

\subsection{Description in gravity}\Label{reviewgrav}

By analyzing symmetries, the half-BPS states described above should be dual in the semiclassical limit to solutions of $IIB$ supergravity with $SO(4) \times SO(4) \times U(1)$ symmetry with 5-form flux and constant dilaton.  These solutions have been found in \cite{llm}:
\begin{equation}
ds^2 = -h^{-2} \, (dt+V_idx^i)^2 + h^2 \, (dy^2 + dx^idx^i) + R^2 \,
d\Omega_3^2 + \tilde{R}^2 \, d\tilde{\Omega}_3^2,
\Label{llmmetric}
\end{equation}
where the coefficients are given in terms of a function $u(x^1,x^2,y)$ as
\begin{equation}
R^2 = y \, \sqrt{\frac{1-u}{u}}, \quad \tilde{R}^2 = y \,
\sqrt{\frac{u}{1-u}}, \quad h^{-2} = \frac{y}{\sqrt{u(1-u)}}.
\Label{llmfunctions1}
\end{equation}
The one form $V$ is
\begin{equation}
V_i(x_1,x_2,y) = -\frac{\epsilon_{ij}}{\pi} \int_{\mathbb{R}^2}
\frac{u(x_1',x_2',0) \, (x_j-x_j') \,   \, dx_1'dx_2'}{[(\vec{x}-\vec{x}')^2
+y^2]^2} \, .
\Label{llmfunctions2}
\end{equation}
Thus, the function $u$ completely specifies the solution.   This function in turn satisfies a harmonic equation in $y$ and as such is fully determined by its boundary condition in the $y=0$ plane
\begin{equation}
u(r,\varphi,y) = \frac{y^2}{\pi} \int_{\mathbb{R}}
\frac{u(r',\varphi',0) \,\,  d^2\vec{r}'}{[(\vec{r}-\vec{r}')^2+y^2]^2} \, ,
\Label{llmufunction}
\end{equation}
where we have used polar coordinates for the $(x^1,x^2)$ plane.    The full solution also contains a 5-form field strength that we are omitting here.

A dictionary between these solutions and the half-BPS field theory states has been  established in \cite{llm,vijayjoan}.  The $(x_1,x_2)$ plane at $y=0$ is identified with the single particle oscillator phase space in the dual gauge theory:
\begin{equation}
(x_1,x_2) \leftrightarrow (p,q) \, .
\label{phasespaceident}
\end{equation}
The Planck scale $l_p$ is related to $\hbar$ in the field theory:
\begin{equation}
l_p^4 \leftrightarrow \hbar \, .
\Label{lplanck}
\end{equation}
Finally, the boundary condition function $u(x^1,x^2,0)$ is identified with the single particle phase space distribution in the fermionic description of half-BPS states in the field theory.    The eigenstates of $M_k$ that we are interested in are all rotationally invariant in phase space, so, taking
\begin{equation}
r^2 = (x^1)^2 + (x^2)^2 \leftrightarrow p^2 +q^2
\end{equation}
and writing  $u(x^1,x^2,0)$ as $u(r)$ we could identify
\begin{equation}
u(r) \leftrightarrow 2\pi \hbar W(r), \ 2\pi \hbar \textrm{Hu}(r),
\ 2\pi g(r)
 \Label{udef}
 \end{equation}
We will see in Sec.~\ref{classmoments} that  a classical observer will not be able to detect the differences between these choices.

\section{Integrability and the gravitational solution}
\Label{sec:genmulti}

Above we saw that a half BPS state in field theory with a fixed set of
fermion excitation energies $\CF$ can be completely identified by a measurement of a set of gauge-invariant observables, the  moments $\CM$.   However, the gravitational description of half BPS states involves the  effective single particle phase space distribution (\ref{udef}) and as such appears to lose information about the $N$ fermion excitation energies that are necessary to characterize the state completely.    Here we show that for eigenstates of the moment operators $M_k$, no information is lost.   All the moments of the fermion energies are stored in angular moments of the gravitational solution that can be measured from infinity.  In this way the gravitational solution preserves the integrable character of half BPS states.

\subsection{Multipole expansion and integrable charges}
\Label{sec:multipole}

At infinity, a natural tool for studying the state of a spacetime is
the analysis of the multipole moments of all fields.   Such an expansion is usually done at the level of observable quantities, but since all half BPS geometries are characterized by the scalar function $u(x^1,x^2,y)$, it suffices to study the asymptotic multipoles of $u$.  We choose the boundary condition for $u$ on the $y=0$ plane as corresponding to the Wigner distribution on the fermion phase space of the dual field theory:
\begin{equation}
u(r) = 2\pi \hbar W(r)
\Label{uisWigner}
\end{equation}
 following (\ref{udef}). Then, recalling the exact expression (\ref{eq:wigner}) for the Wigner distribution associated to a half-BPS eigenstate of the $M_k$, we expand the denominator of (\ref{llmufunction}) as a power series in $\frac{r'^2 - 2\vec{r}\cdot \vec{r}'}{r^2+y^2}$.  The integrals can be done explicitly using \begin{eqnarray}
\int_0^{\infty} \int_0^{2\pi} dr  \, d\varphi \, e^{-r^2} L_f(2r^2) & r^{l+1} & (n_1 \cos \varphi + n_2 \sin \varphi)^p \\
& = & \frac{\pi f!}{2^p} \binom{p}{\frac{p}{2}} \sum_{k=0}^f
\frac{(-1)^k 2^k (k+\frac{l}{2})!}{(f-k)!(k!)^2}
\delta_{p,\textrm{even}}\, , \nonumber
\end{eqnarray}
where $|\vec{n}|^2=1$.    Spacelike infinity is reached in the solutions (\ref{llmmetric}) by either going to large radial distances in the $(x^1,x^2)$ plane at $y=0$ or by going to large y.  Thus it is natural, at infinity, to introduce the new radial coordinate $\rho$:
\begin{equation}
r = \rho \sin \theta, \quad y = \rho \cos \theta, \quad
\textrm{with } \theta \in [0,\frac{\pi}{2}] \, .
\Label{newradial}
\end{equation}
In this coordinate system the boundary of the spacetime lies at
$\rho \to \infty$. The new angular variable $\theta$, which
measures the angle between the two radial variables $\{r,\,y\}$
becomes the azimuthal angle in the 5-sphere of the asymptotic
$AdS_5\times S^5$ geometry.

After some manipulation we find that in this coordinate system $u$
has the asymptotic multipole expansion
\begin{equation}
u(\rho,\,\theta)
=  2\cos^2 \theta \sum_{l=0}^{\infty}
\frac{\hbar^{l+1}\sum_{f\in \mathcal{F}} A^l(f)}{\rho^{2l+2}}
(-1)^l(l+1)\, \, {_2F_1}(-l,l+2,1;\sin^2 \theta),
\Label{eq:expansion}
\end{equation}
%\begin{eqnarray}
%u(\rho,\theta,\varphi) & = & 2\cos^2 \theta \sum_{l=0}^{\infty}
%\frac{\hbar^{l+1}\sum_{f\in \mathcal{F}} A^l(f)}{\rho^{2l+2}}
%\sum_{m=0}^l \binom{m+l}{2m} \binom{2m}{m} (-1)^{m+l} (m+l+1)
%\sin^{2m} \theta \nonumber \\
%& = & 2\cos^2 \theta \sum_{l=0}^{\infty}
%\frac{\hbar^{l+1}\sum_{f\in \mathcal{F}} A^l(f)}{\rho^{2l+2}}
%(-1)^l(l+1)\, \, {_2F_1}(-l,l+2,1;\sin^2 \theta),
%\Label{eq:expansion}
%\end{eqnarray}
where ${_2F_1}$ is the hypergeometric function and $A^l(f)$ is a
polynomial of order $l$ in $f$:
\begin{equation}
A^n(f) \equiv \sum_{s=0}^f \frac{(-1)^{f-s}2^s f!}{(f-s)!s!}
(s+1)_n \, .
\end{equation}
The Pochhammer symbol is defined by $(\alpha)_n = \alpha
(\alpha+1)\cdots (\alpha+n-1) =
\frac{(\alpha+n-1)!}{(\alpha-1)!}$.   Here $\hbar \leftrightarrow l_p^4$ as in (\ref{lplanck}). All of these sums can be computed by introducing the generating function
\begin{equation}
B(n,f,a) = \sum_{s=0}^f \binom{f}{s} (-1)^{f-s} a^{s+n} =
a^n(a-1)^f\,,
\end{equation}
from which we can derive
\begin{equation}
A^n(f) = \left( \frac{d}{da} \right)^n B(n,f,a) \arrowvert_{a\to
2}.
\end{equation}
The first few sums are
\begin{eqnarray}
A^0(f) & = & 1 \, , \nonumber \\
A^1(f) & = & 2f+1 \, , \nonumber \\
A^2(f) & = & (2f + 1)^2 + 1 \, , \nonumber \\
A^3(f) & = & (2f + 1)^3 + 5 (2f + 1) \, , \nonumber \\
A^4(f) & = & (2f + 1)^4 + 14 (2f + 1)^2 + 9 \, , \nonumber \\
A^5(f) & = & (2f + 1)^5 + 30 (2f + 1)^3 + 89 (2f + 1)\,, \nonumber
\end{eqnarray}
These are even and odd in $(2f + 1)$ as can be 
seen from the recurrence relation
\begin{equation}
A^{n+1}(f) = (2f+1) A^{n}(f) + n^2 A^{n-1}(f)\,,
\end{equation}
that allows us to compute any such sum.   Another convenient form is
\begin{equation}
A^n(f) = \left( \frac{d}{da} \right)^n B(n,f,a) \arrowvert_{a\to
2} =  n! \sum_{k=0}^n \binom{n}{k} \binom{f}{k} 2^k \, .
\end{equation}

Since the whole ten dimensional geometry is determined by the
function $u$, (\ref{eq:expansion}) permits computation of all the angular moments of the metric.   This is given in Appendix~\ref{app:metric}.   For example, the asymptotic multipole expansion of the $V_1$ component of the $V_i$ shift vector in (\ref{llmmetric}) is
\begin{eqnarray}
V_1(r,\varphi,\theta) & = & - 2 \sin \varphi \sin \theta \left(
\frac{\hbar N}{\rho^3}+ \sum_{l=1}^{\infty}
\frac{\hbar^{l+1}\sum_{f\in
\mathcal{F}} A^l(f)}{\rho^{2l+3}} (-1)^l(l+1) \right. \cdot \nonumber \\
& \cdot & \left[ \, {_2F_1}(-l,l+2,1;\sin^2 \theta) + l \, \,
{_2F_1}(1-l,l+2,2;\sin^2 \theta)\right] \bigg).
\end{eqnarray}
The angular moments of curvature invariants can be computed from this data.

\paragraph{Measuring the charges:}
In this multipole expansion, the data about the underlying state $\{ \mathcal{F}\}$
enters the l$^{{\rm th}}$ moment in sums of the form
\begin{equation}
\sum_{f\in \mathcal{F}} A^l(f) = \sum_{k=0}^l c_k \, M_k
\end{equation}
where the $M_k$ are the moments defined in (\ref{momentdef}) and $c_k$ is the coefficient of $f^k$ in the polynomial expansion of $A^l(f)$.   Thus a measurement of the first $N$ multipole moments of the metric functions can be inverted to give the set of charges $\CM$ of the underlying state, from which the complete wavefunction can be reconstructed.

\subsection{The semiclassical limit}
\Label{classmoments}

The semiclassical limit for half BPS states is
\begin{equation}
\hbar \to 0 ~~~;~~~~ \hbar N = \alpha = \textrm{fixed} \, .
\Label{classlim}
\end{equation}
Since the moments $M_l$ scale with $N$ as
\begin{equation}
M_l = m_l N^{l+1}
\Label{Mscaling}
\end{equation}
we see that as $\hbar \to 0$
\begin{equation}
    \hbar^{l + 1} M_l \to m_l \,  \alpha^{l+1} \, .
\end{equation}
Thus, in the semiclassical limit
the multipole expansion reduces to
\begin{equation}
u(\rho,\,\theta) = 2\cos^2 \theta \sum_{l=0}^{\infty}
\frac{2^l\alpha^{l+1}m_l}{\rho^{2l+2}} \, (-1)^l(l+1)\, \,
{_2F_1}(-l,l+2,1;\sin^2 \theta).
 \Label{eq:wignerlimit}
\end{equation}
The l$^{\rm th}$ multipole is dominated by $M_l$ with subleading corrections from the lower moments.
Thus, in the semiclassical limit,  the amplitude of each multipole moment  is directly related to a higher order conserved charge of the underlying integrable system.

The computations above were carried out by identifying the boundary condition for the $u$ function in the $y=0$ plane with the Wigner distribution in the harmonic oscillator phase space (\ref{uisWigner}).   However, there are other distributions on phase space which differ in the ordering prescription that they assume for quantum mechanical operators.   Any of these distributions is a candidate for identification with the $u$ function and it was proposed in \cite{vijayjoan} and each choice should be appropriate for describing the effective metric sensed by different quantum gravity observables.   However, since ordering prescriptions only lead to differences in observables at higher orders in $\hbar$, the leading semiclassical result (\ref{eq:wignerlimit}) should be universal.   We demonstrate this below for the Husimi distribution (\ref{husimi}) and for the grayscale distribution (\ref{grayscale}).

\paragraph{Husimi distribution:}  We can carry out the multipole expansion of Sec.~\ref{sec:multipole} after identifying the boundary condition for the $u$ function on the $y=0$ plane as
\begin{equation}
u(r) = 2\pi\hbar \, \textrm{Hu}(r)
\end{equation}
where the Husimi distribution $\textrm{Hu}(r)$ for a given get of fermion excitation energies $\CF$ is given in (\ref{husimi}).  We obtain
\begin{equation}
u_H(\rho,\,\theta) = 2\cos^2 \theta \sum_{l=0}^{\infty}
\frac{2^l \hbar^{l+1}\sum_{f\in \mathcal{F}}
\frac{(f+l)!}{f!}}{\rho^{2l+2}} \, (-1)^l(l+1)\, \,
{_2F_1}(-l,l+2,1;\sin^2 \theta). \Label{eq:husimi}
\end{equation}
Since $\frac{(f+l)!}{f!} = f^l + \textrm{(lower powers of $f$)}$,
the semiclassical limit gives
\begin{equation}
u_H^{s.c.} \to 2\cos^2 \theta \sum_{l=0}^{\infty} \frac{2^l
\alpha^{l+1} m_l}{\rho^{2l+2}} \, (-1)^l(l+1)\, \,
{_2F_1}(-l,l+2,1;\sin^2 \theta). \Label{eq:husimilimit}
\end{equation}
This agrees exactly with the semiclassical limit in
(\ref{eq:wignerlimit}).   Thus the semiclassical asymptotic observer will always measure the same multipole moments.  However, the subleading terms that are suppressed in powers of $\hbar$ (or, equivalently, $l_p$, see (\ref{lplanck})) differ between the geometries based on the Wigner and Husimi distributions.  Presumably, this implies that while the classical spacetime does not depend on the operator ordering prescription,  the underlying ``quantum foam''  \cite{vijayjoan} looks different to
different quantum mechanical observables.

\paragraph{Grayscale distribution: } Finally we can identify the boundary condition for $u$ with the grayscale distribution (\ref{grayscale}) as
\begin{equation}
u(r) = 2\pi \, g(r) \, .
\end{equation}
In this case it is diffcult to analyze the multipole expansion in complete generality.  Thus we present the computation for a state whose semiclassical limit curve (see (\ref{limit1})) is
\begin{equation}
y(x) = (\delta - 1) x ~~~;~~~  x\in (0,N-1)
\Label{triangle}
\end{equation}
This limit curve corresponds to the extremal superstar spacetime (see \cite{myerstafjord,vijayjoan}  for details).  The corresponding grayscale phase space distribution is (\ref{grayscale})
\begin{equation}
u = \frac{1}{1+y'} = \frac{1}{\delta}, \quad 0 \le r \le r_0 \, ,
\end{equation}
where $r_0 = \sqrt{2\hbar \delta N}$ to leading order in $N$.   Because the curve (\ref{triangle}) is entirely determined by $\delta$, the moments $M_k$ are all expressible in terms of this parameter:
\begin{equation}
M_k = \sum_{i=1}^N f_i^k = \sum_{i=0}^{N-1} (\delta i)^k =
\delta^k \sum_{i=0}^{N-1} i^k \approx \delta^k
\frac{N^{k+1}}{k+1} \, .
\end{equation}
Then we can once again compute the asymptotic expansion and find that in the semiclassical limit
\begin{equation}
u_G =  2\cos^2 \theta \sum_{l=0}^{\infty}
\frac{2^l\alpha^{l+1}m_l}{\rho^{2l+2}} \, (-1)^l(l+1)\, \,
{_2F_1}(-l,l+2,1;\sin^2 \theta) \, ,
\end{equation}
where $m_l = { \delta^l \over l+1}$
This matches with the Wigner (\ref{eq:wignerlimit}) and Husimi (\ref{eq:husimilimit}) geometries as expected.

\section{Semiclassical observers and information loss}

Above we showed that the entire tower of conserved  charges of half BPS states is stored, in the gravitational description, in multipole moments of the spacetime.   We will now show that a semiclassical observer will not be able to access this information.   First, we will argue that such an observer, with access to length scales longer than the Planck length, will only be able to measure the very low multipoles.  Second, we will show that these low multipoles are essentially universal for almost all states. Thus, semiclassical observers will necessarily lose information even though it is present in the full theory.

\subsection{High multipoles cannot be measured}

Recall that the semiclassical limit for half BPS states is $\hbar \to 0$ with $\hbar N = \alpha$ fixed (\ref{classlim}).   Translating this into gravity using $l_p^4 \leftrightarrow \hbar$ this amounts to
\begin{equation}
\hbar N \leftrightarrow l_p^4 N \sim g_s l_s^4 N \sim L^4 = \alpha = {\rm fixed} ~~~~~;~~~~~
L  \sim l_s (g_s N)^{1/4}
\Label{classlim2}
\end{equation}
where $l_s$ is the string length, $g_s$ is the string coupling, and $L$ is the length scale associated to the asymptotic $\ads{5} \times S^5$ spacetime using the standard AdS/CFT dictionary.  Thus, the semiclassical limit (\ref{classlim}) that we have been using is the same as the standard limit in the AdS/CFT correspondence, namely  $g_s \to 0, N \to \infty$ with   $L$ fixed.

One way of measuring the l$^{{\rm th}}$ multipole in
(\ref{eq:expansion},\ref{eq:wignerlimit}) is to compute the 
(2l)$^{{\rm th}}$ derivative of the metric functions or any
suitable invariant constructed from them.     Consider an
apparatus of finite size $\lambda$ that makes such a measurement.
In order to compute the k$^{{\rm th}}$ derivative of a  quantity
within a region of size $\lambda$,  the apparatus will have to
make measurements at a scale $\lambda/k$.      However, a
semiclassical apparatus can only measure quantities over distances
larger than the Planck length.  Thus, the k$^{{\rm th}}$
derivative can only be measured if
\begin{equation}
{\lambda \over k} > l_p = g_s^{1/4} l_s
\Label{bound1}
\end{equation}
Setting the size of the apparatus to be a fixed multiple of the AdS scale
\begin{equation}
\lambda = \gamma L \, ,
\end{equation}
this says that
\begin{equation}
k < \gamma N^{1/4}
\end{equation}
for a derivative to be semiclassically measurable.   In order to
identify the underlying quantum state we have shown that $O(N)$
multipoles must be measured.  Since $N^{1/4}/N \to 0$ as $N \to
\infty$ we see that the semiclassical observer has access to a
negligible fraction of the information needed to identify the
quantum state.

In order to measure the first $N$ multipoles {\it without} making
Planckian measurements, an
observer would require an apparatus of size $\lambda \sim N^{3/4}
L$.  But in the semiclassical limit, this size diverges.    This
is an interesting appearance of the connection in gravity between
the extreme UV and the extreme IR.

We might attempt to avoid the difficulty of probing high order
derivatives in spacetime by directly measuring the angular moments
of (\ref{eq:expansion},\ref{eq:wignerlimit}) along the $\theta$
direction.   Consider a location where this circle has a
circumference $\lambda$.  Then to measure the k$^{{\rm th}}$
multipole, an apparatus will have to measure the amplitude of a
fluctuation along this circle which has $O(k)$ nodes.   For a
semiclassical apparatus, the spacing between nodes must be bigger
than Planck length in order to be measurable.  This requires
$\lambda/k > l_p$ just as in (\ref{bound1}).   Then by reasoning
identical to the above, only $O(N^{1/4})$ multipoles will be
measurable.  One might attempt to increase the number of
measurable mutipoles by moving out to locations at which the
$\theta$ circle has a large circumference.  However, as described
below (\ref{newradial}), $\theta$ is the azimuthal coordinate in
the $S^5$ part of the asymptotic $\ads{5} \times S^5$ in the
geometry.  As such its size remains of $O(L)$ even at infinity,
and direct measurement of higher multipoles remains impossible for
the semiclassical observer.

\subsection{Low multipoles are universal}
\label{pks}

The argument above has shown that almost all information concerning the detailed quantum state is unmeasurable by a semiclassical observer.  This leaves open the possibility that the lowest $N^{1/4}$ multipoles can be measured.   Here we will argue that even these measurements cannot be done with sufficient precision to distinguish between most microstates.

This is because almost all half BPS states of the kind that we are considering lie very close to a ``typical state'' \cite{vijayjoan}.  Thus, as described in Sec.~\ref{reviewgauge}, almost all half BPS states with a bound on individual fermion excitation energies lie close to the curve
\begin{equation}
f(x) = \left ( 1 + {N_c \over N} \right ) \, x \, .
\end{equation}
(Recall the continuum notation (\ref{limit1}).)   Equivalently, this is the typical spectrum of fermion energies in half-BPS states for which $0 \leq f_1 < f_2 < \cdots < f_N \leq N_c + N$.  We will show that the standard deviation to mean ratio of the moments $M_k$ (\ref{momentdef}) is small in such an ensemble of states, implying that the differences between states cannot be observed by a semiclassical observer.

First  define the  ensemble of integers
\begin{equation}
\CE_1 = \{f_i  \, | \, 0\leq f_1 < f_2 < f_3 \cdots < f_N \leq N_c + N \}
\Label{ensemble1}
\end{equation}
The standard deviation to mean ratio of the moments (\ref{momentdef}) 
\begin{equation}
{\sigma(M_k)_{\CE_1} \over \langle M_k \rangle_{\CE_1}}
\end{equation}
is computed by averaging $M_k$ and $M_k^2$ over the ensemble (\ref{ensemble1}), i.e.  $\sigma(M_k)_{\CE_1}^2 = \langle M_k^2 \rangle_{\CE_1} - \langle M_k \rangle_{\CE_1}^2$.  The inequalities between the $f_i$ make these averages cumbersome.   Thus, we define a slightly different ensemble
\begin{equation}
\CE_2 = \{f_i  \, | \, 0\leq f_1 \leq f_2 \leq f_3 \cdots \leq f_N \leq N_c + N \}
\Label{ensemble2}
\end{equation}
in which the integers are allowed to be equal.  Averages in this ensemble will turn out to be easier to compute.    First we demonstrate the bound
\begin{equation}
{\sigma(M_k)_{\CE_1} \over \langle M_k \rangle_{\CE_1}} 
<
{\sigma(M_k)_{\CE_2} \over \langle M_k \rangle_{\CE_2}}  \, .
\Label{bound}
\end{equation}
To show this bound consider the auxiliary integral
\begin{equation}
I = N (N + N_c)^k \int_0^1 x^k \, dx \, .
\Label{integral}
\end{equation}
Sums of the form $\sum_i f_i^k$ with $f_i$ drawn from either $\CE_1$ or $\CE_2$ can be considered as  discrete approximations to this integral.   In the $N \to \infty$ limit, an average over either $\CE_1$ or $\CE_2$ will converge to (\ref{integral}) via the results in the standard Monte Carlo theory of computing integrals (see Sec.~3 of \cite{montecarlo}).   Thus, as $N \to \infty$ the mean of $M_k$ taken in either ensemble will be the same.  Now observe that every set of $f_i$ that appears in the ensemble (\ref{ensemble1}) also appears in  (\ref{ensemble2}).  However, (\ref{ensemble2}) contains additional sets of integers in which some $f_i$ coincide.  But sets of coinciding $f_i$ provide poorer approximations to the integral (\ref{integral}).   Thus the variance of $M_k$ computing in the second ensemble (\ref{ensemble2}) must be larger.  This shows the bound (\ref{bound}).

In the large $N$ continuum limit, expectation values in the ensemble (\ref{ensemble2}) can be computed by turning the sums over integers into integrals.\footnote{The reader may wonder why the same continuum approximation does not apply to the ensemble (\ref{ensemble1}), obviating the need for the bound (\ref{bound}).   When $N_c \gg N$, averages in (\ref{ensemble1}) can indeed be approximated in this way.   However, when $N_c$ is $O(N)$ the continuum limit is more subtle and  (\ref{bound}) is necessary.}     We should find that $\langle 1 \rangle_{\CE_2} = 1$ and to this end we compute the normalization constant
\begin{equation}
C = \sum_{r_N=0}^{N_c + N} \sum_{r_{N-1}=0}^{r_N} \cdots
\sum_{r_1=0}^{r_2} 1 \approx \int_0^{N_c + N} dr_N \int_0^{r_N}
dr_{N-1} \ldots \int_0^{r_2} dr_1 = \frac{(N_c + N)^N}{N!} \, .
\end{equation}
When $f(r_1,\ldots,r_N)$ is symmetric in its
arguments, we have the useful identity
\begin{equation}
\int_0^{N_c + N} dr_N \int_0^{r_N} dr_{N-1} \ldots \int_0^{r_2} dr_1 f
= \frac{1}{N!} \int_0^{N_c + N } dr_N \int_0^{N_c + N} dr_{N-1} \ldots
\int_0^{N_c + N} dr_1 f.
\end{equation}
Using this we can compute the mean
\begin{equation}
\langle M_k \rangle_{\CE_2} = \frac{1}{C}\int_0^{N_c + N} dr_N \int_0^{r_N}
dr_{N-1} \ldots \int_0^{r_2} dr_1 M_k = \frac{N(N_c + N)^k}{k+1}
\end{equation}
It is straightforward to compute $\langle M_k^2 \rangle_{\CE_2}$
similarly, and this gives the standard deviation to mean ratio as
\begin{equation}
\label{stdevtomean}
\frac{\sigma(M_k)_{\CE_2}}{\langle M_k \rangle_{\CE_2}}  = \frac{k}{\sqrt{N(2k+1)}}.
\end{equation}
This vanishes for small $k$, and is of order one when
$k \sim N$.  Using the bound (\ref{bound}) we can conclude that in the semiclassical $N \to \infty$ limit, almost all half BPS states have essentially identical low moments.    Therefore the corresponding classical solutions have essentially identical low order multipoles, and the differences will not be observable by a semiclassical observer.

\section{Discussion}
\Label{sec:discussion}

We showed that the multipole expansion of half BPS asymptotically
$\ads{5} \times S^5$ spacetimes  encodes the tower of  commuting,
conserved charges that completely identifies the underlying quantum
eigenstate.   We then argued that a semiclassical observer, having
access to coarse-grained observables, would only be able to
measure the low multipoles, and that these were essentially
universal for almost all half-BPS states.   Thus
the semiclassical observer necessarily loses information.   The
basis of the latter argument was that features of spacetime that
occur at the Planck  scale are only accessible in quantum gravity,
and not to classical observers.  This raises a question as to why
the multipole expansion of the spacetime metric that we studied is
itself reliable at high orders.   In fact, strictly speaking, it
is not classically reliable at very high orders -- in quantum
gravity there is a wavefunction over metrics that will lead to
significant fluctuations in the precise form of the highly
suppressed higher order multipoles.  However, because we are
constructing spacetimes dual to exact eigenstates of the moment
operators, upon quantization the eigenvalues should still be
extractable from the wavefunction of spacetime.

\section*{Acknowledgements}
We thank Jan de Boer,  Tamaz Brelidze, Veronika Hubeny, Vishnu Jejjala, Thomas Levi, Don Marolf, Rob Myers, Mukund Rangamani, Simon Ross, Slava Rychkov and Jung-Tay Yee  for useful discussions. B.C.,  V.B.\ and J.S.\ are supported in part by the DOE under grant DE-FG02-95ER40893, by the NSF under grant PHY-0331728 and by an NSF Focused Research Grant DMS0139799. K.L.\ is supported by the Netter Fellowship from the University of Pennsylvania.  B.C. would like to dedicate this paper to his parents, Bo{\.{z}}ena St{\c{e}}pnik-{\'{S}}wi{\c{a}}tek, Kip Sumner, and David Reichman.

\appendix
\section{Asymptotic form of the 10D metric}
\Label{app:metric}

Having computed the asymptotic expansion of the function $u$ as in Sec.~\ref{sec:multipole}, one can find the multipole expansion of every component of the metric.  The complete metric is given in (\ref{llmmetric},\ref{llmfunctions1},\ref{llmfunctions2},\ref{llmufunction}).    A multipole expansion for $V_i$ (\ref{llmfunctions2}) can be calculated as was done for $u$ in section \ref{sec:multipole}. The answer is \begin{eqnarray}
V_1(r,\varphi,\theta) & = & - 2 \sin \varphi \sin \theta \left(
\frac{\hbar N}{\rho^3}+ \sum_{l=1}^{\infty}
\frac{\hbar^{l+1}\sum_{f\in
\mathcal{F}} A^l(f)}{\rho^{2l+3}} (-1)^l(l+1) \right. \cdot \Label{eq:Vexpansion} \\
& \cdot & \left[ \, {_2F_1}(-l,l+2,1;\sin^2 \theta) + l \, \,
{_2F_1}(1-l,l+2,2;\sin^2 \theta)\right] \bigg). \nonumber
\end{eqnarray}
The second component is given by $V_2 = -\cot \varphi V_1$. With
these tools it is now possible to compute the components of the
metric to any  order.  The first terms of $V_i$ are
\begin{eqnarray}
V_1 & = & -\frac{2\sin \varphi \sin \theta}{\rho^3} \left( \hbar N - \frac{2\hbar^2(2-3\sin^2 \theta)(2E+N)}{\rho^2} \right. \nonumber  \\
& & + \left. \frac{6\hbar^3 (3-12\sin^2 \theta +10\sin^4 \theta)(2M_2 + 2E + N)}{\rho^4} + \mathcal{O}(\frac{1}{\rho^6}) \right),   \\
V_2 &=& - \cot\varphi \, V_1
\end{eqnarray}
The first terms in the scalar functions in the metric are
\begin{eqnarray}
R^2 & = &   \frac{\rho^2}{\sqrt{2\hbar N}} \left( 1 + \hbar \frac{(2\frac{E}{N} + 1)(1-3\sin^2 \theta) - \cos^2 \theta N}{\rho^2}
+ \mathcal{O}(\frac{1}{\rho^4})\right), \nonumber \\
\tilde{R}^2 & = &  \sqrt{2\hbar N}\cos^2 \theta \left( 1 - \hbar\frac{(2\frac{E}{N} + 1)(1-3\sin^2 \theta) -
\cos^2 \theta N}{\rho^2} + \mathcal{O}(\frac{1}{\rho^4}) \right), \nonumber \\
h^{-2} & = & R^2 + \tilde{R}^2 = \frac{\rho^2}{\sqrt{2\hbar N}} \left( 1+ \hbar\frac{\cos^2 \theta + (2 \frac{E}{N} +1)(1-3\sin^2 \theta)}{\rho^2}
+ \mathcal{O}(\frac{1}{\rho^4}) \right), \nonumber \\
h^{2} & = & \frac{\sqrt{2\hbar N}}{\rho^2} \left( 1-
\hbar\frac{\cos^2 \theta + (2 \frac{E}{N} +1)(1-3\sin^2
\theta)}{\rho^2} + \mathcal{O}(\frac{1}{\rho^4}) \right).
\nonumber
\end{eqnarray}
These quantities are more directly observable than the function
$u$, but due to their cumbersome nature it is easier to work with
$u$.  Taking the semiclassical limit as in
Sec.~\ref{classmoments}, we again see that at each order of the
expansion a new moment $M_i$ appears, as was expected from the
behavior of $u$.

\end{document}